\newcommand{\Proof}{\noindent{\em Proof: }}
\newcommand{\QED}{\hskip 10pt plus 1filll $\Box$ \vspace{10pt}}

\newtheorem{theorem}{Theorem}
\newtheorem{proposition}[theorem]{Proposition}
\newtheorem{lemma}[theorem]{Lemma}

\newtheorem{definition}[theorem]{Definition}

\documentclass{article}

\usepackage{amssymb,amsmath,epsf}
\usepackage{wasysym}

\begin{document}

\title{An Algebraic Topological Construct of Classical Loop Gravity and the prospect of Higher Dimensions }        
\author{ Madhavan Venkatesh \\ \\  Centre For Fundamental Research and Creative Education\\ Bangalore, India} 
       
\date{May 2, 2013}
\maketitle

\begin{abstract}
In this paper, classical gravity is reformulated in terms of loops, via an algebraic topological approach. The main component is the loop group, whose elements consist of pairs of cobordant loops. A Chas-Sullivan product is described on the cobordism, and three other products, namely the 'vertical', 'horizontal' and 'total' products are re-introduced. (They have already been defined in an earlier paper by the author). A loop calculus is introduced on the space of loops, consisting of the loop variation functional,the loop derivative, Mandelstam derivative, and what the author wishes to call, the Gambini-Pullin contact functional. The loop derivative happens to be a generator of the group of loops, and the Gambini-Pullin functional is an infinitesimal generator of diffeomorphisms. A toy model of gravity is formulated in terms of the above, and it is proven that the total product provides for the Maurer-Cartan structure of the space. Further, new quantities labeled as 'momenta', 'velocity' and 'energy' are introduced in terms of the loop products and loop derivatives. The prospect of re-constructing general relativity in higher dimensions is explored, with the 'momenta' and 'velocity' being the basic loop variables. In 
the process, it is proven that the 'momenta' and 'velocity' behave as cobordant loops in higher dimensions.

\end{abstract}
\section{Introduction and Overview of the new loop program}
This paper initiates the new loop program that will encompass the subjects talked of, below. Forthcoming papers will be based on the higher dimensional loop representation of gravity ,the possible quantization approaches and the quantum theory itself
The aim of this formalism is, at its ends, to construct a quantum theory of gravity and provide for a step towards unification. First, a 5+1 split of gravity is done, and the constraints are linearized in terms of the new loop variables. Then, the Master Constraint is introduced and the principal constraint that is, $\mathrm{Q}$, which corresponds to the 'energy' in ordinary dimensions;(arises due to more degrees of freedom in higher dimensions) is packed into it. With the classical theory completely reformulated (in the Ashtekar approach), prospects of quantization will be explored including loop quantization, twistor, deformation and Berezin-Toeplitz quantization. In a pre-quantum set-up; it is found that the Grassmannian of the Hilbert Space is actually a Calabi-Yau manifold, with a $\mathit{U(1)}$ compactification, analogous to Kaluza-Klein theory. This is particularly interesting as the theory will incorporate supersymmetry into it. In order to recover the physics of the $\mathit{U(1)}$, a detour into surgery theory will be made, involving a handle decomposition, using the fact that the compactified piece is K\"{a}hler with an exact curvature two-form. As for the quantum states, polynomials in knots are used; as they are both Cobordism and diffeomorphism invariant and prove to be ideal for the scenario.\\
So, these developments in prospect are the motivation for the new loop approach.

\subsection{Preliminaries}

We begin with a compact Lie Group $\mathrm{G}$ and the Lie algebra associated with it $\mathfrak{
g}$. We have the commutation relation $\left[ \mathfrak{g},\mathfrak{g}\right] =\hat{\mathfrak{g
}}$. The loop group $\mathrm{LG}$ is the group of maps from the circle $\mathrm{S^{1}}$ into $
\mathrm{G}$, with $\hat{\mathfrak{g}}$ being the loop algebra endowed with it, and $\Omega\mathrm
{G}$ the loop space. We restrict the elements of $\mathrm{LG}$ to be ordered pairs of disjoint 
cobordant loops. We endow with $\Omega\mathrm{G}$ a loop homology $\mathbb{H}$ and a 
Chas-Sullivan type product,$\circ$ that, along with the action of the 'loop bracket' $
\left\lbrace\centerdot,\centerdot\right\rbrace $ makes $\mathbb{H}$ into a Gerstenhaber algebra. 
It can easily be shown that the brackets follow the Jacobi identity. We 
now restrict the Killing form of $\mathfrak{g}$ to its Cartan sub-algebra, thus, inducing a 
metric on the Lie group manifold $\mathcal{M}$.
We have, the Maurer-Cartan structure equation $$d\phi=-\frac{1}{2}\left[ \phi,\phi\right], $$ 
where $\phi$ is the Maurer-Cartan 1-form; that also behaves as the connection form,or, in the 
language of Reiris-Spallanzani \cite{RS}, it is the universal one-form.
(Note: This one-form is the one whose curvature happens to be the loop derivative introduced by 
Gambini-Pullin. This is an extremely useful tool to construct a (loop) polynomial action, which 
can be varied, using the loop derivative and the connection derivative to give the equations of 
motion, or, more precisely, the vacuum Einstein equations;in another form).

\subsection{The Loop Products}
in this section, we give a description of the three products defined in    , the 
"vertical","horizontal" and "total" products. We deal with two free, analytic loops $
\alpha$ and $\beta$ to express the products between them. First, we define a quantity known as 
the holonomy average,(which will be used to define the products themselves) $$\gamma=\frac{1}{2}
\left\lbrace \left( \oint dx^{a}dx^{b}\phi_{ab}\right) +\left( \oint dy^{a}dy^{b}\phi_{ab}\right
) \right\rbrace $$.

\begin{definition}
We define the 'vertical product' as the following composition:
$$\alpha\oplus\beta=\left(\alpha\circ\beta\right)\circ\gamma,$$
\end{definition}
where the $\circ$ is the Chas-Sullivan loop product, that forms a graded, commutative 
associative algebra. 
\begin{definition}
 The "horizontal product" is written as : $$\alpha\ominus\beta=\left(\alpha\circ\gamma\right)+
\left(\beta\circ
\gamma\right).$$
\end{definition}

\begin{definition}  Further, we write the 'total product' as a composition of the two; 
 $$\alpha\circledast\beta=\left\lbrace \left( \alpha\oplus\beta\right) \circ\left( 
\alpha\ominus\beta\right) \right\rbrace. $$
 \end{definition}
\begin{proposition}
The integral over all space of the total product $\circledast$ between two loops, $\alpha$ and $
\beta$ can be represented as the curvature of the universal connection form, $d\phi$.
\end{proposition}
We leave the reader to ponder over this, while a formal and rather indirect proof will be 
presented after introduction to some loop calculus, that is, the three derivatives; the loop 
derivative, the Mandelstam derivative and the Gambini-Pullin variation (or connection functional).
\begin{definition}
We define the loop variation functional $\alpha$ with path  $\varrho^{x}_{0} \circ \varrho^{0}_{x
}$ by basing its point of conjoiment to another point $x+\epsilon u$, with the composition of an 
infinetisimal vector at that point, $\delta u$, which is also tangential to the original path of 
the loop. That is,
\begin{equation}
\delta\alpha=\varrho^{x}_{0}\circ\delta u\circ\varrho^{0}_{x+\epsilon u}
\end{equation}
\end{definition}

\begin{definition}
We now define the loop derivative $\Delta_{ab}$ acting on the loop $\alpha$ as follows:
\begin{equation}
\Delta_{ab}\left( \alpha^{x}_{0}\right) = \partial_{a}\delta_{b}\left( x\right) -\partial_{b}\delta_{a}
\left( x\right) + \left[ \delta_{a}\left( x\right) ,\delta_{b}\left( x\right) \right] 
\end{equation}
\end{definition}

\begin{definition}
The connection contact functional, or the Gambini-Pullin variation, is defined for a loop, with respect 
to a connection $\phi$:

\begin{equation}
\frac{\delta\alpha}{\delta\phi^{a}\left( x\right) }=\oint_{\alpha}dx^{b}\delta\left( y-x\right) 
\Delta_{ab}\left( \alpha^{x}_{0}\right) \phi\left( x\right).
\end{equation}
\end{definition}

\begin{definition}
We introduce what is called the Mandelstam derivative. Here, it acts like how a covariant 
derivative does for Gauge theories:
\begin{equation}
\mathit{D_{a}}\alpha\left( \varrho^{x}_{0}\right) =\partial_{a}\alpha\left( x\right) +\mathit{i}
\phi_{a}\left( x\right) \alpha\left( x\right) 
\end{equation}
\end{definition}
\subsection{Proof of Proposition 4, and further notes}
\Proof
\
Proposition 4 states: $$\int_{\Omega\mathrm{G}}\alpha\circledast\beta=d\phi$$

First, we consider the space of smooth maps $C^{\infty}\left( S^{1},\mathrm{G}\right) $, on a 
Lie group $\mathrm{G}$ with respect to a specific inner product. Naturally, the loop algebra $
\hat{\mathfrak{g}}$ is the tangent space and the space of sections of the pullback bundle $P\left
( TG\right) $.

We consider the connection to be constructed compatible with the  quantization condition: 
$$\phi_{ab}=\phi_{a}\phi_{b}-\phi_{b}\phi_{a} + \phi_{\left[ a,b\right]} $$
We define the curvature operator, depending on the Sobolev space parameter $s$, as:
$$\Theta^{s}\left( \alpha,\beta\right) =\phi^{s}_{\alpha}\phi^{s}_{\beta}-\phi^{s}_{\beta}\phi^{s
}_{\alpha}+\phi^{s}_{\left[ \alpha,\beta\right]}. $$
Here, the parameter $s$ is $1/2$, as we are dealing with a flat connection.
Finally, we define the inner product on $LG$ , for two loops $\alpha$ and $\beta$ as follows:
$$<\alpha,\beta>=\left\lbrace \left( 1+\Delta\right) ^{s} \left( \alpha\circledast\beta\right) 
\right\rbrace ,$$
where $\Delta$ is the Chas-Sullivan operator, that, along with the composition $\circ$ and the 
"loop brackets " $\left\lbrace \cdot,\cdot\right\rbrace $ make the homology $\mathbb{H}$ into a 
Gerstenhaber qua Batalin-Vilkovisky algebra. It appears, as if the inner-product defined, is a 
little vague, but; the product $\circledast$ appears in the term, giving the complete structure 
of the group $LG$ to the inner product. So, the definition of the inner product in terms of the 
"total product " is justified.
Now, we exploit the K\"{a}hler structure of the loop group to further the proof.
Let $LG$ be the group of loops and $\Omega\mathrm{G}$ be the loop space. The curvature operator 
and the Chas-Sullivan product are defined with respect to the cobordism $\partial W=
\alpha\sqcup\beta$. In this notation, the symplectic form on $LG$ is given by \begin{equation}
\omega\left( \alpha,\beta\right) =\int_{\Omega\mathrm{G}}<\alpha,\beta>,
\end{equation}
when an arbitrary loop $\zeta\in\hat{\mathfrak{g}}$ can be written as the following Fourier 
series,
$$\zeta\left( e^{2\pi it}\right) =\sum_{n>0}\left( \zeta_{n}e^{2\pi i nt}-i\zeta_{n}e^{-2\pi i n 
t}\right) .$$
Here, the symplectic form is just the curvature 2-form, which; can as well be written as $d
\phi$. 
Expanding the inner product, we have,
\begin{equation}
\omega\left( \alpha,\beta\right) =d\phi=\int_{\Omega\mathrm{G}}\left( 1+\Delta \right)^{s} \left( 
\alpha\circledast\beta\right) 
\end{equation}

The  $\left( 1+\Delta\right) ^{s}$ becomes trivial for real exponents(as $s=1/2$, which 
indicates a flat connection). So, neglecting that term, the proof is complete.
\QED

Note: This theorem allows us to plug in different products between loops, vary them (that is, by 
a loop functional or derivative); and arrive at the equations of motion.

For example, let us consider an action,
$$ S\left( \alpha,\beta\right) = \int\left\lbrace \left( \alpha\oplus\beta\right) +\left( 
\alpha\ominus\beta\right) \right\rbrace \sqrt{g} d^{3}x,$$ where $g$ is the metric induced by 
restricting the Killing form $\kappa$ to a Cartan sub-algebra of $\mathfrak{g}$. The equations 
of motion for such a set-up can easily be derived by varying the action with respect to the 
loops(i.e. acting the loop functional $\delta$, defined in Definition 5, on it).
Quite clearly, this is equivalent to the Einstein-Hilbert formalism. We vary the action (with respect to the loops) to obtain the Ricci-flat equations of motion, (by acting the loop variation functional) as follows:
\begin{equation}
$$ $$
\delta S=\int\left\lbrace \left[ \left( \varrho^{x}_{0}\circ\delta u\circ\varrho^{0}_{x+\epsilon_{1} u}\right) \circ\beta\circ\gamma\right] +\left[ \alpha\circ\left( \varrho^{y}_{0}\circ\delta v\circ\varrho^{0}_{y+\epsilon_{2} v}\right) \circ\gamma\right] \right\rbrace 
\end{equation}

$$+ \int\left[ \left\lbrace \left( \varrho^{x}_{0}\circ\delta u\circ\varrho^{0}_{x+ \epsilon_{1} u}\right) \circ\gamma\right\rbrace +\left\lbrace \left( \varrho^{y}_{0}\circ\delta v\circ\varrho^{0}_{y+\epsilon_{2} v}\right) \circ\gamma\right\rbrace \right] $$

$$-\frac{1}{2}\int\left[ \left\lbrace \left( \alpha\circ\beta\circ\gamma\right) \right\rbrace +\left\lbrace \left( \alpha\circ\gamma\right) +\left( \beta\circ\gamma\right) \right\rbrace \right] \left( g_{ab}\delta g^{ab}\right) =0.
$$

where the loops $\alpha$ and $\beta$ have a variation in their paths from points $x$ to $x+\epsilon_{1}u$ and $y$ to $y+\epsilon_{2}v$ respectively. We denote, by $\varrho$, the path; and by $\delta u$ and $\delta v$, infinitesimal vectors at points $u$ and $v$ respectively.

 Let us go a little bit 
deeper, into this interesting set-up. We derive the canonical momenta, $\tilde{\pi}$, by varying 
the action with respect to the connection as follows:

$$
 \tilde{\pi}=\frac{\delta}{\delta\phi}\left\lbrace \left( \int \alpha\oplus\beta\right) +\left( 
\int \alpha\ominus\beta\right) \right\rbrace 
$$

$$
=\left[ \left\lbrace \int\left( \frac{\delta\alpha}{\delta\phi}\circ\beta\circ\gamma\right) +\left( \alpha\circ\frac{\delta\beta}{\delta\phi}\circ\gamma\right) \right\rbrace +\left\lbrace \int\left( \frac{\delta\alpha}{\delta\phi}\circ\gamma\right) + \int\left( \frac{\delta\beta}{\delta\phi}\circ\gamma\right) \right\rbrace \right] 
$$

\begin{equation}
= \left\lbrace \int\left( \left( \oint_{\alpha}dx^{b} \delta \left( y-x\right) \Delta_{ab}\left( \alpha^{x}_{0}\right) \phi^{a} \right) \circ\beta\circ\gamma\right) +\left( \alpha\circ\left( \oint_{\beta}dy^{b}\delta\left( x-y\right) \Delta_{ab}\left( \beta^{y}_{0}\right) \phi^{a}\right) \circ\gamma\right) \right\rbrace \end{equation}
 $$+\left\lbrace \int \left( \left( \oint_{\alpha}dx^{b}\delta\left( y-x\right) \Delta_{ab}\left( \alpha^{x}_{0}\right) \phi^{a}\right) \circ\gamma\right) + \int\left( \left( \oint_{\beta}dy^{b}\delta\left( x-y\right) \Delta_{ab}\left( \beta^{y}_{0}\right) \phi^{a}\right) \circ\gamma\right) \right\rbrace  $$

We now define a quantity, which we shall call 'velocity' for brevity, as:
\begin{equation}
\varpi=\int\mathfrak{i_{\mathrm{X}}}\left\lbrace \int\left( \alpha\oplus\beta\right) +\int\left( \alpha\ominus\beta\right) \right\rbrace,
\end{equation}
where the $\mathfrak{i_{\mathrm{X}}}$ is the interior derivative, with respect to a Hamiltonian vector field, that satisfies the symplectic equation $\mathfrak{i_{\mathrm{X}}}\omega=d\mathrm{H}$.
Further, we write another quantity, called 'Energy'($\mathrm{Q}$); as a total product between the momenta and 'velocity' as follows:
\begin{equation}
\mathrm{Q}=\int\left(\tilde{\pi}\circledast\varpi\right).
\end{equation}

Note: Equipped with this definition, we can now write the structure equation for another space in terms of the above total product, which means, the 'energy' $\mathrm{Q}$ for this space corresponds to the structure of another, wherein; our fundamental loops would be $\tilde{\pi}$ and $\varpi$, and not $\alpha$ and $\beta$. In order to do so, we look for a space wherein $\tilde{\pi}$ and $\varpi$ behave as fundamental loops. The answer to this, is that we move to higher dimensions. In ordinary dimensions (that is, 4) $\tilde{\pi}$ and $\varpi$ are the momenta and 'velocity' respectively. When we move to a higher dimension ( $D+2=6$), we choose our loops to be $\tilde{\pi}$ and $\varpi$; and have them behave the same way as the ordinary loop variables, here, do.

\section{Higher dimensional Gravity}

In this section, we attempt to validate the claim that the variables $\tilde{\pi}$ and $\varpi$ can be used, or, behave as loops in higher dimensions (in particular, 6). 
The products introduced in the previous section hold for our new loop variables $\tilde{\pi}$
 and $\varpi$, generated by the loop algebra $\left[ \mathfrak{g},\mathfrak{g}\right] =\Omega\mathfrak{g}$. We denote by $LG$ the group of analytic loops whose elements consist of ordered pairs of cobordant loops, two of which, are $\tilde{\pi}$ and $\varpi$.
We choose higher dimensions, or, in particular, 6 dimensions for our momenta $\tilde{\pi}$ and 
the velocity $\varpi$ to behave as loops. We need them as observables for a higher dimensional 
theory, and can be sure about the fact that it would reproduce the same theory, in lower 
dimensions, as, it can easily be seen, that; the total product, not only provides for the 
structure of a theory, but also the dynamics of another, in a different dimension!.

\subsection{Some theorems}
\begin{lemma}
The loops $\tilde{\pi}$ and $\varpi$ are cobordant in 6 dimensions.
\end{lemma}
\Proof
Let $LG$ be the group of loops whose elements are ordered pairs of cobordant loops. Let $\tilde{\pi}\in LG$ and $\varpi \in LG $. We need to prove that $\tilde{\pi}$ and $\varpi$ are cobordant. 
 We denote the loop algebra, generating the loops as $\Omega \mathfrak{g}$, arising from the Lie algeebra $\mathfrak{g}$ with Lie Group $\mathrm{G}$ and Group manifold $\mathcal{M}$,
which is an oriented Riemannian six-fold. Cobordisms are defined with respect to loops on $\Omega \mathrm{G}$, the loop space.  It is known that cobordisms are equivalence relations. It is enough if we show that cobordisms of $\tilde{\pi}$ and $\varpi$ 's cobordism class forms an equivalence class on $LG$. 
We observe the construct of the variables $\tilde{\pi}$ and $\varpi$. We see that, the polynomial in loops in the two, are the same; but the actual term differs by the operator acting on them. In the case of the former it is the Gambini-Pullin functional , taken with respect to a flat connection $\phi$; whereas in the latter, it is just the interior derivative taken with respect to a Hamiltonian vector field $\mathrm{X}$.
Now, if we prove that the loops $\tilde{\pi}$ and $\varpi$ are diffeomorphic, we get a result that is a direct consequence of the above lemma. Also, we get an other interesting theorem stating that: 
( Note: we put a direct proof for this on hold, and this theorem shall be proved as a direct result of following theorem)
\begin{theorem}
To every loop $\zeta \in LG$, there exists one and only one unique non-trivial diffeomorphic loop, and they form an ordered pair in $LG$, ie. they are cobordant.
\end{theorem}
It is easier to prove this in our case as $\tilde{\pi}$ contains the Gambini-Pullin functional, which is an infinitesimal generator of diffeomorphisms.
Also, if we prove that the two loops are isotopic, then,it would mean they are diffeomorphic; which in turn, implies that they are cobordant.
Let us examine the terms in $\tilde{\pi}$ and $\varpi$. $\tilde{\pi}$ has a Gambini-Pullin functional that gets rid of terms unrelated to the connection. The initial expression $\left( \int \alpha \oplus \beta + \int \alpha \ominus\beta\right) $ behaves as a 2-form (due to the total product, being a two-form), resulting in a 1-form after the operation. In the case of $\varpi$, we have the interior product acting on the same intitial expression; and it is quite clear that the resultant expression here too, is a 1-form (as $\mathfrak{i_{\mathrm{X}}}$ maps p-forms to (p-1) forms). 
We write the two loops as $\tilde{\pi}\left( \Sigma_{1},\mathcal{M_{\mathrm{1}}}\right) $ and $\varpi\left( \Sigma_{2},\mathcal{M_{\mathrm{2}}}\right)$ , where the $\Sigma$s are two surfaces and $\mathcal{M_{\mathrm{1}}}$ and $\mathcal{M_{\mathrm{2}}}$ are two submanifolds of $\mathcal{M}$.
We need to show that if there is an embedding $\Sigma_{1}\amalg\left( -\Sigma_{2}\right) \hookrightarrow \partial W$, then; under the induced embedding: $\left( \Sigma_{1}\times\mathbb{R}\right) \coprod\left( \Sigma_{2}\times\mathbb{R}\right)\hookrightarrow\partial W\times\mathbb{R}$, $\mathcal{M_{\mathrm{1}}}\coprod\left( -\mathcal{M_{\mathrm{2}}}\right) $, bounds an oriented annulus in $\mathcal{M}\times\mathbb{R}$. We can prove this via the Whitney embedding theorem. The above embeddings result in two points with opposite intersection numbers, say $x$ and $y$. Now, by the embedding theorem; there exists an isotopy on $\mathcal{M}$, that is constant in a neighbourhood of $\mathcal{M_{\mathrm{1}}}\cap \mathcal{M_{\mathrm{2}}} - \left\lbrace x,y\right\rbrace $;\

$f: \mathcal{M}\rightarrow\mathcal{M_{\mathrm{1}}}\cap\mathcal{M_{\mathrm{2}}}$. It is also apparent, that there is an induced structure on $\mathcal{M} \times \mathbb{R}$, that is an $\mathit{S^{1}}\times\left( 0,1\right) $ embedding; that is topologically equivalent to the annulus. This proves, that the two loops are isotopic, therefore, cobordant and diffeomorphic.
\QED
\section{Conclusion}
In this paper, we have initiated some theorems and supplemented them with proofs; that will be helpful to construct a theory of quantum general relativity, prospects of which have been put forward in the 'Introduction' section. These topological constructions go on to be extremely useful; when building a physical theory. This paper initiates the foundations of the program that will eventually lead to an algebraic topological formalism of quantum gravity. It is intended to use the usual techniques of 'Loop Quantum Gravity'; to the classical theory of general relativity (ie. the Ashtekar variables) as well as exploring prospects of quantization; that include loop quantization, twistor; and Berezin-Toeplitz quantization in forthcoming papers.

\section*{Acknowledgements}
I would like to thank Martin Reiris and Jorge Pullin for a clarification.
This work was carried out at the Center For Fundamental Research And Creative
Education (CFRCE), Bangalore, India, under the guidance of Dr B S Ramachandra
whom I wish to acknowledge.  I would like to acknowledge the Director Ms. Pratiti B R for creating the highly charged research atmosphere at CFRCE. I would also like to thank my fellow researchers Magnona H Shastry,Vasudev Shyam, Karthik T Vasu and Arvind Dudi.


\begin{thebibliography}{8}

\bibitem{MV} Venkatesh, Madhavan.
\textit{A Construct of Dynamics, Space and Gravity from Loops}. arXiv:1211.1131

\bibitem{PS} Pressley, Andrew , Segal, Graeme. \textit{Loop Groups}. Oxford Mathematical Monographs. 1986.
\bibitem{GP} Gambini, Rodolfo , Pullin, Jorge.
\textit{Loops, Knots, Gauge theories and Quantum Gravity}. Cambridge University Press. 2000.

\bibitem{RS} Reiris, Martin, Spallanzani, Pablo.
\textit{The loop derivative as a curvature}. arXiv:math/9802080



\bibitem{T} Turaev, Vladimir
\textit{Cobordism of knots on surfaces}.Journal of Topology, Oxford University Press. 2008.

\bibitem{CS} Chas, Moira , Sullivan, Dennis
\textit{String Topology} arXiv:math/9911159

\bibitem{S} Salamon, Dietmar
\textit{Notes on flat connections and the loop group} Citeseer. 1998.



\bibitem{Q} Larra{\'{i}}n-Hubach, Andr{\'{e}}s
\textit{The order of curvature operators on loop groups}. Letters in Mathematical Physics. Springer. 2009.
  



\end{thebibliography}
\end{document}